\documentclass[%
 reprint,%
 amssymb, amsmath,%
 aip,cha,%
]{revtex4-1}

\usepackage{graphics}
\usepackage{amsmath}
\usepackage{amsfonts}
\usepackage{latexsym}
\usepackage{amsbsy}
\usepackage{epstopdf}
\usepackage{bm}%

\newcommand{\ket}[1]{\vert #1 \rangle}

\usepackage{color}

\usepackage{ulem} 

\newcommand{\dyadic}[1]{{#1}
\setbox0=\hbox{$\scriptstyle\leftrightarrow$}
   \setbox2=\hbox{$#1$}
   \dimen0=.5\wd0 \advance\dimen0 by-.5\wd2
   \advance\dimen0 by-\wd0
   \kern\dimen0
{^{\hbox{$\scriptstyle\leftrightarrow$}}}}

\begin{document}

\title{Quantum Physics Meets Music: A ``Real-Time'' Guitar Recording Using Rydberg-Atoms and Electromagnetically Induced Transparency}
\thanks{Publication of the U.S. government, not subject to U.S. copyright.}
\author{Christopher~L.~Holloway}
\email{christopher.holloway@nist.gov}
\author{Matthew T. Simons}
\author{Abdulaziz H. Haddab}
\affiliation{National Institute of Standards and Technology, Boulder,~CO~80305, USA}
\author{Carl J. Williams}
\affiliation{National Institute of Standards and Technology, Gaithersburg,~MD,~20899,~USA}
\author{Maxwell W. Holloway}
\affiliation{Cooperative Institute for Research in Environmental Sciences, Boulder,~CO~80305, USA}

\date{\today}

\begin{abstract}

We demonstrate how Rydberg atoms and the phenomena of electromagnetically induced transparency can be used to aid in the recording of a musical instrument in real time as it is played. Also, by using two different atomic species (cesium and rubidium) in the same vapor cell, we demonstrate the ability to record two guitars simultaneously, where each atomic species detects and allows for the recording of each guitar separately. The approach shows how audio data (the musical composition) can be detected with a quantum system, illustrating that due to the research over the past decade we can now control ensembles of atoms to such an extent that we can use them in this ``entertaining'' example of recording a musical instrument.



\end{abstract}

\maketitle



Rydberg atoms are atoms with one or more electrons excited to a very high principal quantum number $n$ \cite{gal}. These atoms have several useful properties that scale as $n$. They have very large dipole moments (that scale as $n^2$). Their polarizability scales as $n^7$, and their lifetime scales as $n^3$. The spacing between the Rydberg levels scales as $1/n^3$. Rydberg atoms have large range interactions between each other that scales as $n^4/R^3$ (where $R$ is the  inter-atomic distance) and have a van der Waals interaction that scales as $n^{11}/R^6$. These various properties allow for a large array of applications and interesting physics. For example, (1) the large dipole moments make them sensitive to electric fields, making for good field sensors, (2) the long lifetimes could lead to the development of new laser sources, and (3) the large interaction lengths create the possibilities for qubits and highly-entangled cluster states, just to mention a few.

Significant progress has been made in the development of radio frequency (RF) electric (E) field strength and power metrology techniques based on the large dipole moments associated with Rydberg states of alkali atomic vapor placed in glass cells \cite{r2, r3, r4, r5, r6, r7, r8, r9, r10, r11, r12, r13, r14, r15, r16, r17, r18, r19, r20}.  In this approach, the concept of electromagnetically induced transparency (EIT) is used for the E-field sensing, performed either when the RF field is on-resonance of a Rydberg transition (using Autler-Townes (AT) splitting) or off-resonance (using AC Stark shifts).

This Rydberg-atom based sensor can act as compact reciever/antenna, enabling quantum-based receivers to be used in communication applications to detect and receive modulated signals\cite{dan, rc1, rc2, amfmstereo, biterror, rc3, qpsk, rc4}. This has led to the new term ``atom-radio'' \cite{rc3, atomradio}. Recently we extended the atom receiver to develop a Rydberg atom-based mixer that allows for the measurement of the phase of an RF wave\cite{phase}, which was the needed missing link for Rydberg atom-based quantum sensors to be able to fully characterize the RF E-field in one compact vapor cell.

In this paper we illustrate how this Rydberg-atom EIT-based approach can be used as a means to both record (in real time) the output of a guitar (or any other musical instrument), and to listen to the output of a guitar as it is played through a set of speakers.  We also demonstrate the ability to detect and record two guitars simultaneously by using two different atomic species in the same vapor cell. In this approach, we use the output of the guitar to amplitude modulate (AM) a continuous wave (CW) carrier. The AM modulated carrier is detected and received with the Rydberg atoms. The output of the Rydberg-atom based detector is both played through a set of speakers and is also recorded through a computer.  One way of explaining how this works is, in effect, one is hearing the change in the quantum state of the Rydberg atoms at audio frequencies through the speakers.

\begin{figure}[!t]
\centering
\scalebox{.40}{\includegraphics*{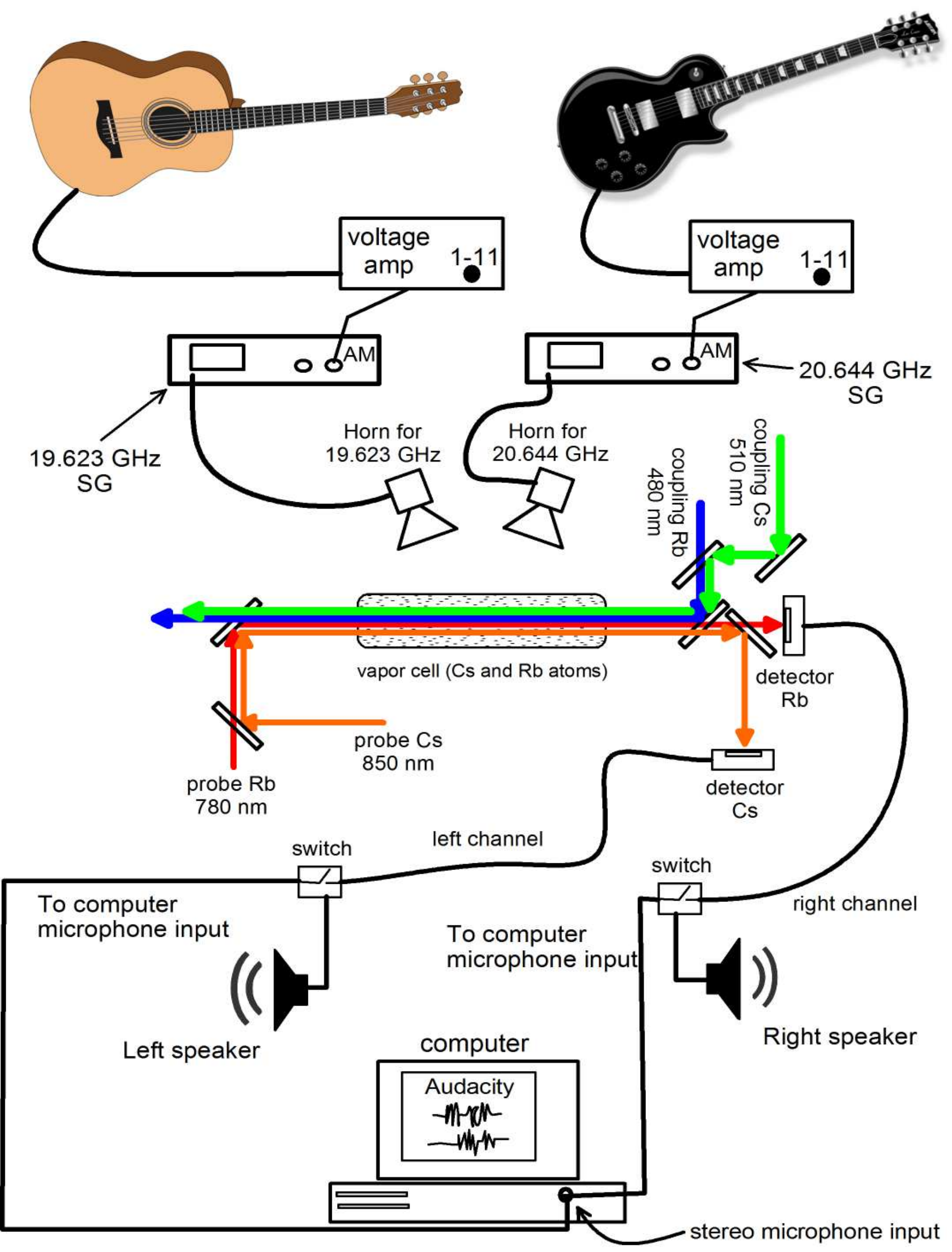}}
\caption{Illustration of a four-level system, the vapor cell setup for measuring EIT, with counter-propagating probe and coupling beams, and experimental setup used to record the output of a guitar.}
\label{setup2}
\end{figure}

The measurement approach is explained in detail in Refs\cite{r2, r3, r4, r5, r6, r7, r8, r9, r10, r11, r12, r13, r14, r15, r16, r17, r18, r19, r20}; here we give a brief explanation. This technique is easily explained by a four-level atomic system illuminated by a single weak (``probe") light field. One laser is used to probe the response of the atoms and a second laser is used to excite the atoms to a Rydberg state (the “coupling” laser), see Fig~\ref{setup2}. [Note that Fig.~\ref{setup2} shows four lasers.  We use two different atomic species (cesium ($^{133}$Cs) and rubidium ($^{85}$Rb)) to simultaneously record two guitars, which required the four lasers, two for $^{133}$Cs (a 850~nm probe laser and a 510~nm coupling laser) and two for $^{85}$Rb (a 780~nm probe laser and a 480~nm coupling laser). Each atomic species works in the same way, the only difference is the wavelengths of the lasers needed for the different two atomic species. The following discussion concentrates on Rb.] In the presence of the coupling laser, the atoms become transparent to the probe laser transmission (this is the concept of EIT). Applying an RF field to this four-level atomic system causes a splitting of the transmission spectrum (the EIT signal) for the probe laser.  This splitting of the probe laser spectrum is easily measured and is directly proportional to the applied RF E-field amplitude (through Planck's constant and the dipole moment of the atom). By measuring this splitting ($\Delta f_m$), we get a direct measurement of the magnitude of the RF E-field strength for a time-harmonic field from \cite{r3, r4}:
\begin{equation}
|E| = 2 \pi \frac{\hbar}{\wp} D\, \Delta f_m= 2 \pi \frac{\hbar}{\wp}\Delta f_0 \,\,\, ,
	\label{mage2}
\end{equation}
where $\hbar$ is Planck's constant (defined in November 2018 in the re-definition of the International System of Units), $\wp$ is the atomic dipole moment of the RF transition (see Ref. \cite{r3, r17} for discussion on determining $\wp$ and values for various atomic states), $\Delta f_m$ is the measured splitting, $\Delta f_o=D\, \Delta f_m$, and $D$ is a parameter whose value depends on which of the two lasers is scanned during the measurement. If the probe laser is scanned, $D=\frac{\lambda_p}{\lambda_c}$, where $\lambda_p$ and $\lambda_c$ are the wavelengths of the probe and coupling laser, respectively. This ratio is needed to account for the Doppler mismatch of the probe and coupling lasers \cite{r14, r15}. If the coupling laser is scanned, it is not required to correct for the Doppler mismatch, and $D=1$. The phase of the RF field can also be measured with a modification of the technique (by using a Rydberg atom-based mixer\cite{phase}).


\begin{figure}
\centering
\scalebox{.2}{\includegraphics*{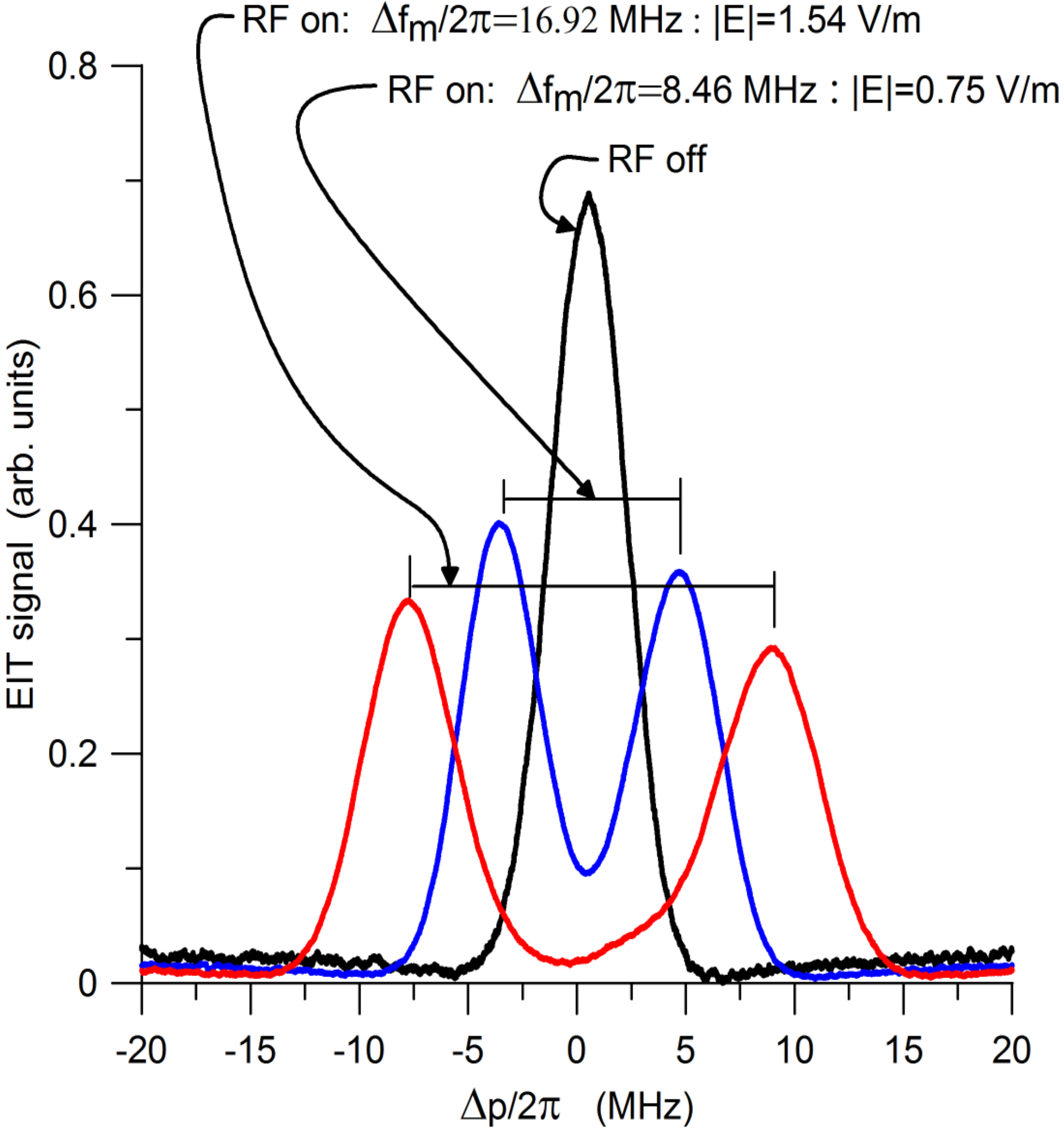}}\hspace{5mm}
\scalebox{.16}{\includegraphics*{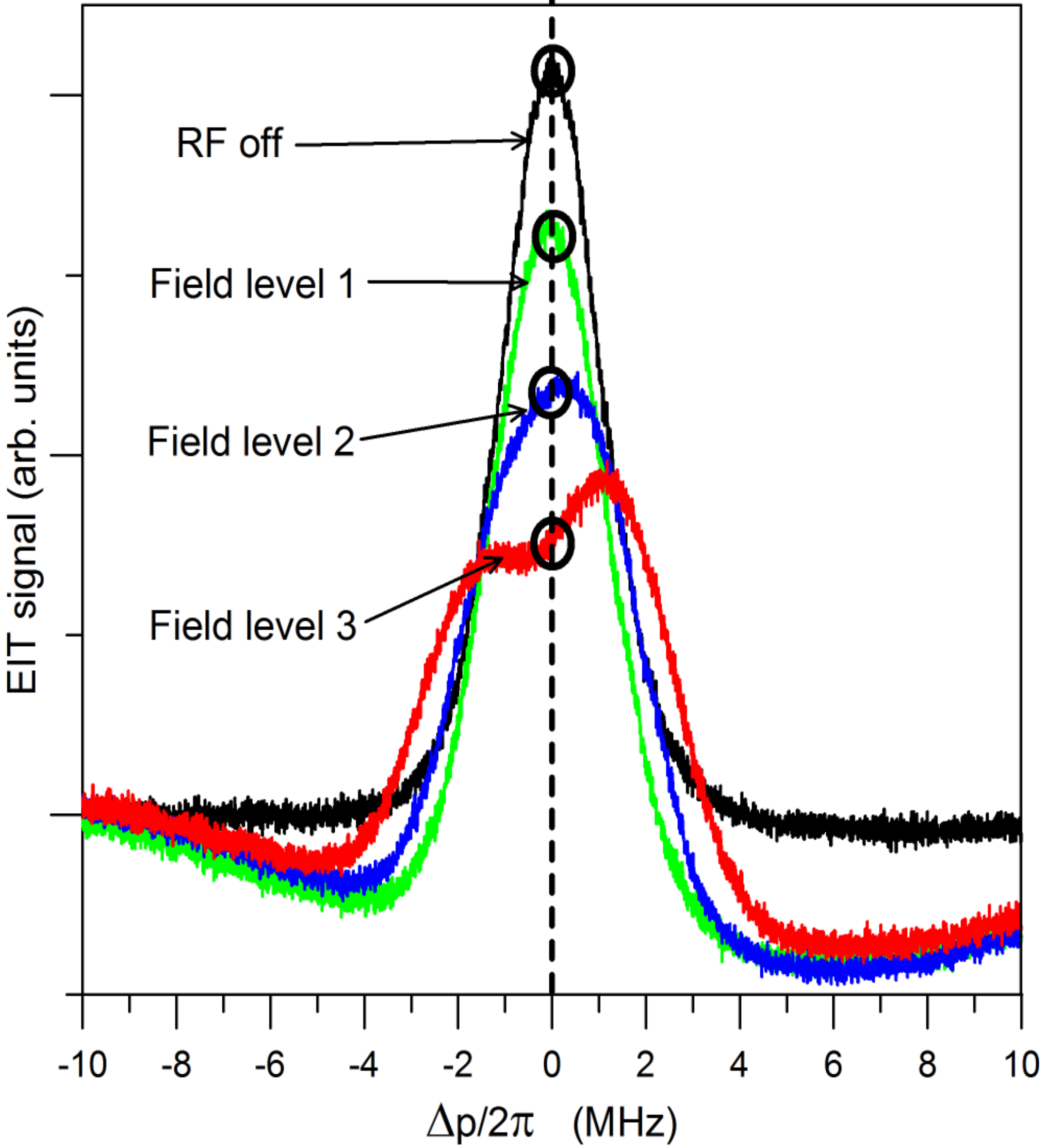}}\\
\vspace*{-2mm}
{\hspace{-1mm}\tiny{(a) \hspace{30mm} (b)}}\\
\caption{Illustration of the EIT signal (i.e., probe laser transmission through the cell) as a function of probe laser detuning $\Delta_p$. This dataset is for 20.64~GHz and corresponds to this following 4-level $^{85}$Rb atomic system: $5S_{1/2}-5P_{3/2}-47D_{5/2}-48P_{3/2}$. (a) splitting of EIT signal\cite{emceurope} and (b) before splitting of EIT signal (AM modulation scheme at the center of the EIT signal).}
\label{eitpeak}
\end{figure}

An example of a measured spectrum for an RF source with different E-field strength is shown in Fig.~\ref{eitpeak}(a). This figure shows the measured EIT signal for two E-field strengths.  These results are for scanning the probe laser, in which $\Delta_p$ is the detuning of the probe laser  (where $\Delta_p=\omega_p-\omega_o$; $\omega_o$ is the on-resonance angular frequency of the Rydberg state transition and $\omega_p$ is the angular frequency of the probe laser). Notice that the AT splitting increases with increasing applied E-field strength. To obtain these results, we use $^{85}$Rb atoms and the levels $\ket{1}$, $\ket{2}$, $\ket{3}$, and $\ket{4}$ correspond respectively to the $^{85}$Rb  $5S_{1/2}$ ground state,  the $5P_{3/2}$ excited state, and two Rydberg states. The coupling laser is locked to the $5P_{3/2}$ -- $47D_{5/2}$ $^{85}$Rb Rydberg transition ($\lambda_c=480.2704$~nm).
The probe is scanned across to the D2 transition ($5S_{1/2}$-$5P_{3/2}$ or wavelength of $780.241$~nm). As is typically done, we modulate the coupling laser amplitude with a 30~kHz square wave and detect any resulting modulation of the probe transmission with a lock-in amplifier. This removes the Doppler background and brings the EIT signal out of the noise, as shown by the curve with the one peak in Fig.~\ref{eitpeak}(a). Application of RF at 20.64~GHz to couples states $47D_{5/2}$ and $48P_{3/2}$ splits the EIT peak (shown by the two other curves in the figure).

There is a minimum RF field level that is required before the splitting shown in Fig.~\ref{eitpeak}(a) occurs.  When an RF-field  strength is increased from zero, the amplitude of the EIT signal decreases and its linewidth broadens before the EIT signal splits into two peaks.  This type of behavior is shown in Fig.~\ref{eitpeak}(b), where results for the EIT signal with no RF  field and three cases for different RF fields strengths are shown.

Here we used amplitude modulation (AM) to detect and record the guitars, in which the output of the musical instrument is used to AM modulate a carrier. We will discuss the situation where the amplitude of the carrier (and the modulation depth) is such that no splitting in the EIT signal will occur (other situations are discussed in Ref.\cite{amfmstereo}). The AM modulated carrier will only cause the peak of the EIT line to move up and down the dashed line shown in Fig.~\ref{eitpeak}(b). Hence, by locking the probe laser to $\Delta_p=0$ (while also locking the coupling laser to $5P_{3/2}$-$47D_{5/2}$ Rydberg transition), the voltage output of the photo-detector (the probe laser transmission) will be directly correlated to the modulating signal.  That is, no demodulation circuity is needed. The Rydberg atoms automatically demodulate the signal and we get a direct read-out of the audio signal (the signal or waveform of the musical composition in our case).

The experimental setup for transmitting, detecting, receiving, and recoding (or listening to) a guitar (or a guitar duet) is shown in Fig.~\ref{setup2}. We used an acoustic guitar with an electronic {\it pickup} and an electric guitar. The voltage output level of each guitar's pickup needed to be amplified before it was used to modulate a CW carrier. The output of the each guitar pickup was fed through an ``in-house`` built voltage amplifier. The voltage amplifier was simply an operational amplifier with one 1~k$\Omega$ resistor and one variable resistor (with maximum value of 100~k$\Omega$, which is controlled by a knob that ranges from 1 to 11). These two voltage waveforms from the two amplifiers were used to AM modulate two different carrier frequencies. The acoustic guitar waveform modulated a 19.626~GHz carrier and the electric guitar waveform modulated a 20.644~GHz carrier.  We used two different signal generators (SG) to generate these two different continuous wave (CW) signals.  The modulation was performed by using the AM feature in the SG. The SG AM feature is limited to 100~kHz, which is adequate since the waveform for audio signals is limited to about 20~kHz. For higher modulation rates, an external mixer is required (as was done for receiving pseudo-random bit streams at different modulation rates\cite{biterror}).


Each SG is connected to two separate standard gain horn antennas and each antenna is placed 15~cm from a cylindrical glass vapor cell containing both $^{133}$Cs and $^{85}$Rb atomic vapor, see Fig.~\ref{setup2}. The vapor cell has a length of 75~mm and a diameter of 25~mm. The $^{133}$Cs atoms are used to receive the 19.626~GHz modulated carrier, and the $^{85}$Rb atoms are used to receive the 20.644~GHz modulated carrier.   Since we use the two atom species in this experiment, four lasers are required, see  Fig.~\ref{setup2}.  The probe laser for $^{133}$Cs is a 850.53~nm laser ($6S_{1/2}$-$6P_{3/2}$) focused to a full-width at half maximum (FWHM) of 425~$\mu$m, with a power of 41.2~$\mu$W. To generate an EIT signal, we couple to the $^{133}$Cs $6P_{3/2}$ -- $34D_{5/2}$ states by applying a counter-propagating coupling laser at $\lambda_c=511.1480$~nm with a power of 48.7~mW, focused to a FWHM of 620~$\mu$m. We apply an RF field at 19.626~GHz to couple states $34D_{5/2}$ and $35P_{3/2}$.  For $^{85}$Rb, the probe laser is a 780.24~nm laser focused to a full-width at half maximum (FWHM) of 400~$\mu$m, with a power of 22.3~$\mu$W. To produce an EIT signal in $^{85}$Rb (using the atomic states given in Fig.~\ref{eitpeak}), we apply a counter-propagating coupling laser (wavelength $\lambda_c=480.271$~nm) with a power of 43.8~mW, focused to a FWHM of 250~$\mu$m.  We apply an RF field at 20.644~GHz to couple states $47D_{5/2}$ and $48P_{3/2}$.

We use two different photodetectors to measure the transmission for each probe laser through the atomic vapor (one for $^{85}$Rb and one for $^{133}$Cs, see Fig.~\ref{setup2}). The output of the photo-detectors (which are voltage waveforms and basically the audio waveforms of the musical composition) are connected in two different arrangements. First, we simply connect the output of the photo-detectors to a set of computer speakers. The output for the $^{133}$Cs probe laser photo-detector was connected to the left computer speaker, and the output for the $^{85}$Rb probe laser photo-detector was connected to the right computer speaker. In the second arrangement, the output of the two photo-detectors are connect to a stereo jack and plugged into the microphone input of a computer. We use the open-source program {\it Audacity} (mentioning this product does not imply an endorsement by NIST, but serves to clarify the software used) to record the guitar composition from the microphone input.

\begin{figure}
\centering
\scalebox{.42}{\includegraphics*{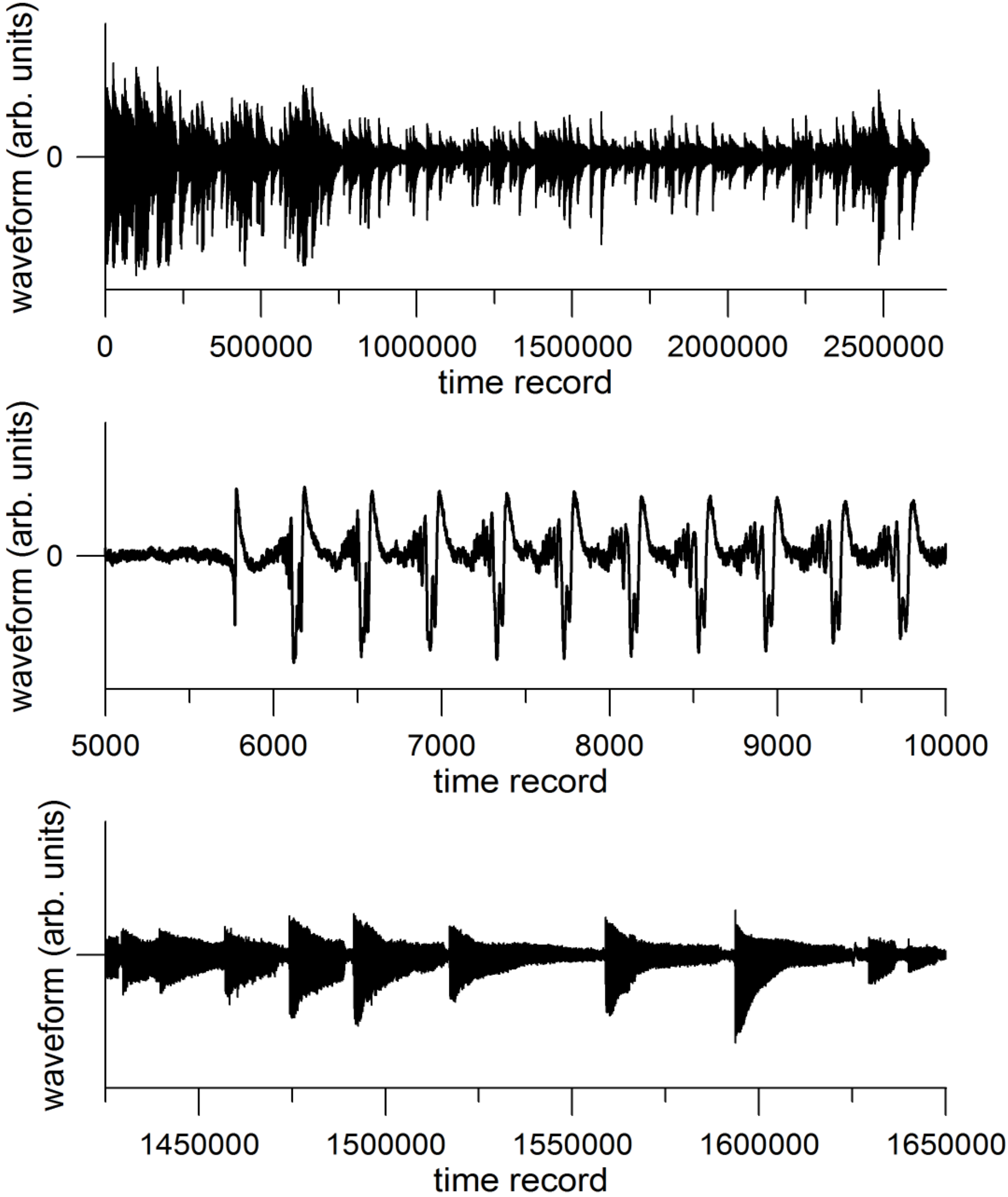}}
\caption{Recorded waveforms: (a) the entire time record (the total time period of these waveforms is approximate 60~seconds),  (b) zoomed in at the start of the guitar recording , and (c) zoomed in in the middle of the guitar recording.}
\label{score}
\end{figure}

We first demonstrate the recording of a single guitar (the acoustic guitar), and before recording the guitar composition, we listened to the guitar through the two speakers.  The output sounds from the speakers (apart from hearing a little noise) was of high fidelity, in that the musical composition was clearly audible and very understandable (see Ref.\cite{amfmstereo} for discussion on audio quantity assessment).  We then recorded the guitar composition (via the microphone input and with {\it Audacity}). The recording is shown in Fig.~\ref{score}.  These curves show the entire recording, as well as zoomed-in part at the beginning, and a zoomed-in part in the middle. This recording was replayed thorough the speakers with high-fidelity and were clearly audible. While noise was present, it had a very minor effect on the quality of the sounds

\begin{figure}
\centering
\scalebox{.42}{\includegraphics*{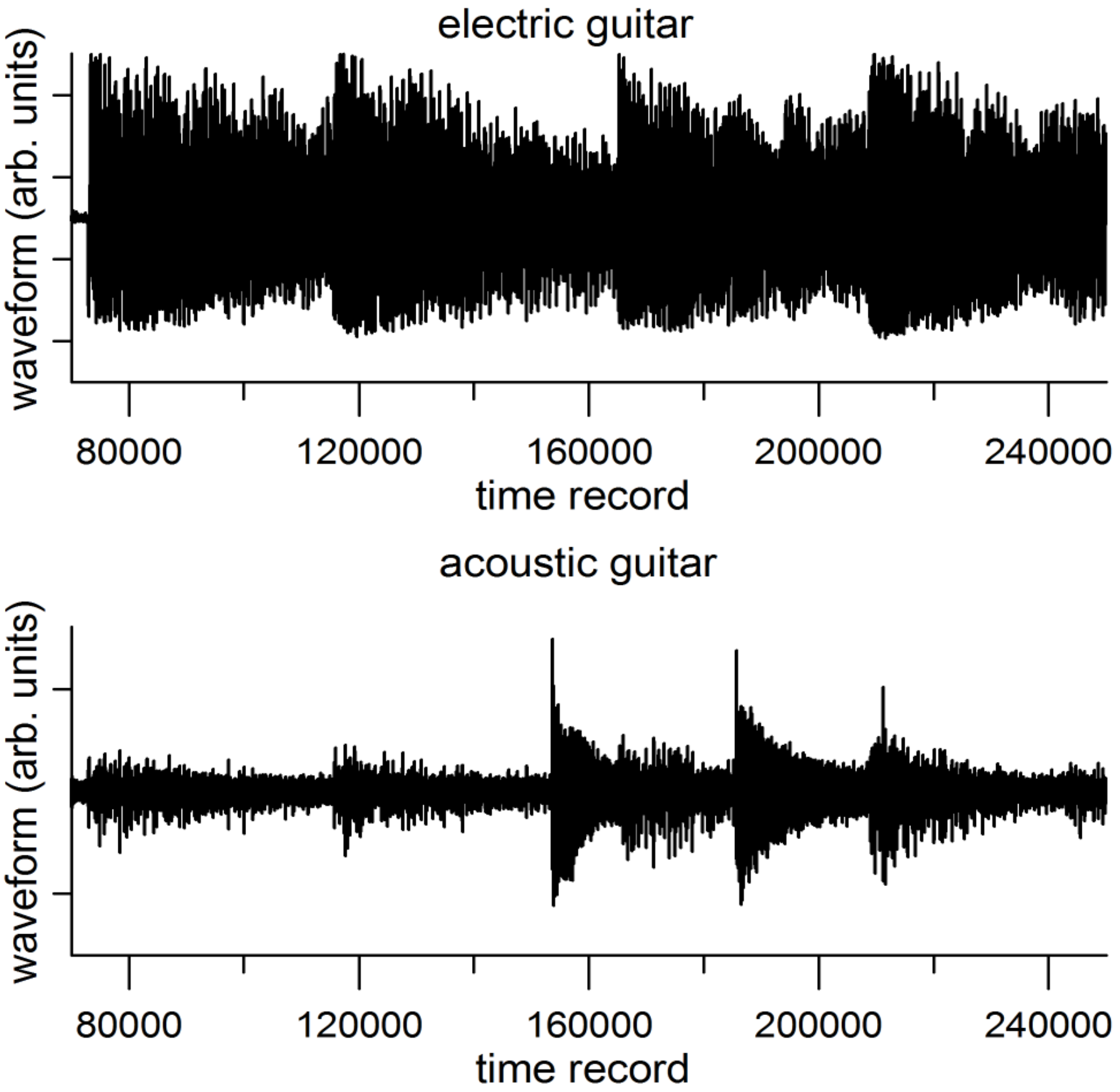}}
\caption{Recorded waveforms from a guitar duet: (a) electric guitar and (b) acoustic guitar (for better visualization, the acoustic guitar data in the figure was put though a high-pass filter such that the notes of the melody can be easily distinguished).}
\label{score2}
\end{figure}

We then played a musical duet with the two guitars.
The output sounds from each speaker (apart from hearing a little noise) was of high fidelity, in that the musical composition was clearly audible and very understandable.  We then recorded the two guitar compositions from each guitar (via the microphone input and with {\it Audacity}). The recordings are shown in Fig.~\ref{score2}.
As before, these recordings were replayed thorough the speakers with high-fidelity and were clearly audible.

In the recording of any musical instrument, one needs to ensure that the recording equipment has enough bandwidth for the data being recorded. A Rydberg-atom receiver has a bandwidth of about 1~MHz to 5~MHz\cite{biterror, qpsk, rc4, phase}.  This bandwidth limit is due to the time required to populate the atoms to a Rydberg state \cite{biterror, qpsk}.  Since music is limited to 20~kHz in frequency, the Rydberg-atom based recorder can capture the full musical range of the instrument with high fidelity.

We demonstrated the ability to use a Rydberg-atom based sensor (or receiver/antenna) to allow us to hear and record in real time a musical composition played on a guitar (or any other type of musical instrument). The simultaneous recording of two guitars was also demonstrated and was accomplished by using two different atomic species placed in the same vapor cell and detecting the EIT signal from each species. While the results in this paper used AM to transit, detect, and record a musical instrument, frequency modulation (FM) of a carrier could also be used (as discussed in Ref.\cite{rc3, amfmstereo}, from the sensing prospective, FM modulation works in a similar way as as AM). In fact, the AM and FM features of a SG have been used to transmit and receive AM and FM radio in stereo\cite{amfmstereo}.

 The results in this work show an interesting applications of ``atomic physics'' applied to the age old topic of musical recording, {\it quantum physics meets music}.  It is quite amazing that over the past decade we have learned to control ensembles of atoms to such an extent that they can be used to record waveforms.   Hopefully, this "entertaining" example of an application of the sometimes esoteric field of quantum physics may entice individuals to study and apply quantum science to a whole new generation of quantum devices and thereby help create the future quantum-based workforce needed to accelerate the field.

\end{document}